\documentclass[11pt,a4paper]{article}
\usepackage{jheppub}
\usepackage{amsmath,amssymb}

\DeclareMathOperator{\tr}{tr} 
\DeclareMathOperator{\Tr}{Tr} 
\DeclareMathOperator{\sgn}{sgn} 
\DeclareMathOperator{\SL}{SL} 
\DeclareMathOperator{\GL}{GL} 
\DeclareMathOperator{\diag}{diag} 
\DeclareMathOperator{\Tor}{Tor} 
\DeclareMathOperator{\Isom}{Isom} 

\title{
Janus configurations with 
$\SL(2,\mathbb{Z})$-duality twists,
Strings on Mapping Tori,
and a Tridiagonal Determinant Formula
}

\author{Ori~J.~Ganor,}
\author{Nathan~P.~Moore,}
\author{Hao-Yu~Sun,}
\author{and Nesty~R.~Torres-Chicon}

\affiliation{
Department of Physics,
  University of California,\\
Berkeley, CA 94720, U.S.A.}

\emailAdd{ganor@berkeley.edu}
\emailAdd{nmoore@berkeley.edu}
\emailAdd{hkdavidsun@berkeley.edu} 
\emailAdd{ntorres@berkeley.edu}

\abstract{
We develop an equivalence between two Hilbert spaces: (i) the space of states of $U(1)^n$ Chern-Simons theory with a certain class of tridiagonal matrices of coupling constants (with corners) on $T^2$; and (ii) the space of ground states of strings on an associated mapping torus with $T^2$ fiber. The equivalence is deduced by studying the space of ground states of $\SL(2,\mathbb{Z})$-twisted circle compactifications of $U(1)$ gauge theory, connected with a Janus configuration, and further compactified on $T^2$.
The equality of dimensions of the two Hilbert spaces (i) and (ii) is equivalent to a known identity on determinants of tridiagonal matrices with corners. 
The equivalence of operator algebras acting on the two Hilbert spaces follows from a relation between the Smith normal form of the Chern-Simons coupling constant matrix and the isometry group of the mapping torus, as well as the torsion part of its first homology group.

}
\subheader{
Preprint:
UCB-PTH-14/04,
}

\begin{document}
\maketitle
\flushbottom

\newcommand{\secref}[1]{\S\ref{#1}}
\newcommand{\figref}[1]{Figure~\ref{#1}}
\newcommand{\appref}[1]{Appendix~\ref{#1}}
\newcommand{\tabref}[1]{Table~\ref{#1}}

\newcommand\rep[1]{{\bf {#1}}} 

\newcommand\SUSY[1]{{${\mathcal{N}}={#1}$}}  
\newcommand\px[1]{{\partial_{#1}}}

\def\be{\begin{equation}}
\def\ee{\end{equation}}
\def\bear{\begin{eqnarray}}
\def\eear{\end{eqnarray}}
\def\nn{\nonumber}

\newcommand\bra[1]{{\left\langle{#1}\right\rvert}} 
\newcommand\ket[1]{{\left\lvert{#1}\right\rangle}} 

\newcommand{\C}{\mathbb{C}}
\newcommand{\R}{\mathbb{R}}
\newcommand{\Z}{\mathbb{Z}}
\newcommand{\Q}{\mathbb{Q}}
\newcommand{\CP}{\mathbb{CP}}

\def\Mst{M_{\text{st}}} 
\def\gst{g_{\text{st}}} 
\def\gIIB{g_{\text{IIB}}} 
\def\lst{\ell_{\text{st}}} 
\def\lP{\ell_{\text{P}}} 

\def\Mf{{\mathbf{M}}} 
\def\lvk{{\mathbf{k}}} 
\def\xR{{R}}
\def\xL{{L}}
\def\gYM{g_{\text{ym}}}

\def\vT{{v}}
\def\Matf{{W}} 
\def\xa{{\mathbf{a}}}
\def\xb{{\mathbf{b}}}
\def\xc{{\mathbf{c}}}
\def\xd{{\mathbf{d}}}

\def\LatCS{{\Lambda}} 
\def\LatMT{{\widetilde{\Lambda}}} 

\def\Id{{\mathbf{I}}} 

\def\xT{{\xi}} 
\def\xI{{\eta}} 

\def\xR{{R}}
\def\xS{{\rho}}

\def\WmId{{H}}

\def\Atd{{\mathcal A}} 
\def\cD{{\mathcal D}} 

\def\Kcpl{{K}} 
\def\Wa{{U}} 
\def\Wb{{V}} 
\def\Ga{{{\mathcal G}_\alpha}} 
\def\Gb{{{\mathcal G}_\beta}} 

\def\Iv{{v}} 
\def\detWmId{{\Delta}} 
\def\IvVec{{\mathbf{v}}}
\def\IuVec{{\mathbf{u}}}

\def\Giso{{{\mathcal G}_{\text{iso}}}} 

\def\opS{{\hat{S}}}
\def\opT{{\hat{T}}}
\def\opM{{\hat{\Matf}}}

\def\hx{{\hat{x}}}
\def\hp{{\hat{p}}}

\def\Jac{{\mathcal J}} 

\def\bPsi{{\overline{\Psi}}}

\def\GWa{{a}} 
\def\GWD{{D}} 
\def\GWe{{\mathbf{e}}} 
\def\GWpsi{{\psi}} 
\def\GWalpha{{\mathbf{\alpha}}} 
\def\GWbeta{{\mathbf{\beta}}} 
\def\GWgamma{{\mathbf{\gamma}}} 
\def\GWr{{\mathbf{r}}} 
\def\GWtdr{{\widetilde{\GWr}}} 
\def\GWu{{\mathbf{u}}} 
\def\GWv{{\mathbf{v}}} 
\def\GWw{{\mathbf{w}}} 
\def\GWq{{\mathbf{q}}} 

\def\sParamEpsilon{{\varepsilon}}

\def\ZLv{{\mathfrak{v}}} 
\def\ZLu{{\mathfrak{u}}} 
\def\OpGa{{{\mathcal O}_{\alpha}}} 
\def\OpGb{{{\mathcal O}_{\beta}}} 
\def\CHuv{{\chi}}

\def\Iso{{\mathcal Y}} 

\def\HChg{{\mathcal R}} 
\def\IuChg{{\tilde{\mathbf u}}} 

\def\Phuv{{\varphi}}

\def\Op{{\mathcal X}} 

\def\tGWpsi{{\tilde{\GWpsi}}} 


\section{Introduction and summary of results}
\label{sec:Intro}

Our goal is to develop tools for studying circle compactifications of \SUSY{4} Super-Yang-Mills theory on $S^1$ with a general $\SL(2,\Z)$-duality twist (also known as a ``duality wall'') inserted at a point on $S^1$.
The low-energy limit of such compactifications encodes information about the operator that realizes the $\SL(2,\Z)$-duality, and can potentially teach us new facts about S-duality itself.
Some previous works on duality walls and related compactifications include \cite{Dabholkar:1998kv}-\cite{Ganor:2012mu}.

In this paper we consider only the abelian gauge group $G=U(1)$, leaving the extension to nonabelian groups for a separate publication \cite{toAppear}.
We focus on the Hilbert space of ground states of the system and study it in two equivalent ways: (i) directly in field theory; and (ii) via a dual type-IIA string theory system (extending the techniques developed in \cite{Ganor:2010md}). As we will show, the equivalence of these two descriptions implies the equivalence of:
\begin{itemize}
\item[(i)]
the Hilbert space of ground states of $U(1)^n$ Chern-Simons theory with action
$$
L = \tfrac{1}{4\pi}\sum_{i=1}^n \lvk_i A_i\wedge dA_i -\tfrac{1}{2\pi}\sum_{i=1}^{n-1} A_i\wedge dA_{i+1}
-\tfrac{1}{2\pi} A_1\wedge dA_n\,,
$$
on $T^2$, and
\item[(ii)]
the Hilbert space of ground states of strings of winding number $w=1$ on a certain target space that contains the {\it mapping torus} with $T^2$ fiber:
$$
\Mf_3\equiv\frac{I\times T^2}{(0,\vT)\sim (1,f(\vT))}\,,\qquad (\vT\in T^2)\,,
$$
where $I=[0,1]$ is the unit interval, and $f$ is a large diffeomorphism of $T^2$ corresponding to the $\SL(2,\Z)$ matrix
\be\label{eqn:SL2Zmatf}
\Matf\equiv
\begin{pmatrix} \lvk_n & -1 \\
1 & 0 \\
\end{pmatrix}
\cdots
\begin{pmatrix} \lvk_2 & -1 \\
1 & 0 \\
\end{pmatrix}
\begin{pmatrix} \lvk_1 & -1 \\
1 & 0 \\
\end{pmatrix}
\,.
\ee
\end{itemize}


We will explain the construction of these Hilbert spaces in detail below.

An immediate consequence of the proposed equivalence of Hilbert spaces (i) and (ii) is the identity
\be\label{eqn:MolinariSpecial}
\det\begin{pmatrix}
\lvk_1 & -1 & 0 &  & -1 \\
-1 & \ddots & \ddots & \ddots & \\
0 & \ddots & \ddots & \ddots & 0 \\
 &\ddots  & \ddots & \ddots & -1 \\
-1 &  & 0 & -1 & \lvk_n \\
\end{pmatrix}
=
\tr\left\lbrack
\begin{pmatrix} \lvk_n & -1 \\
1 & 0 \\
\end{pmatrix}
\cdots
\begin{pmatrix} \lvk_2 & -1 \\
1 & 0 \\
\end{pmatrix}
\begin{pmatrix} \lvk_1 & -1 \\
1 & 0 \\
\end{pmatrix}
\right\rbrack
-2\,.
\ee
which follows from the equality of dimensions of the Hilbert spaces above.
This is a known identity (see for instance \cite{Molinari:2007}), and we will present a proof in \appref{app:DetSmithProof}, for completeness.\footnote{
The continuum limit of \eqref{eqn:MolinariSpecial} 
with $n\rightarrow\infty$ and $\lvk_i\rightarrow 2 + \frac{1}{n^2}V(\frac{i}{n})$ might be more familiar. It leads to a variant of the Gelfand-Yaglom theorem \cite{Gelfand:1959nq} with a periodic potential: $\det[-d^2/dx^2 + V(x)]=\tr\left\lbrack P\exp\oint\begin{pmatrix}
\sqrt{V}+\frac{V'}{2V}  &\quad\frac{V'}{2V} \\
-\sqrt{V} &\quad -\sqrt{V}\\
\end{pmatrix} dx \right\rbrack-2$ (up to a renormalization-dependent multiplicative constant).
}

Moreover,
equivalence of the operator algebras of the systems associated with (i) and (ii) allows us to make a stronger statement.
The operator algebra of (i) is generated by Wilson loops along two fundamental cycles of $T^2$, and keeping only one of these cycles gives a maximal finite abelian subgroup.
Let $\LatCS\subseteq\Z^n$ be the sublattice of $\Z^n$ generated by the columns of the Chern-Simons coupling constant matrix, which appears on the LHS of \eqref{eqn:MolinariSpecial}. Then, the abelian group generated by the maximal commuting set of Wilson loops is isomorphic to $\Z^n/\LatCS$.
The operator algebra of (ii), on the other hand, is constructed by combining the isometry group of $\Mf_3$ with the group of operators that measure the various components of string winding number in $\Mf_3$. The latter is captured algebraically by the Pontryagin dual ${}^\vee(\cdots)$ of the torsion part $\Tor$ of the homology group $H_1(\Mf_3,\Z)$.  (The terms will be explained in more detail in \secref{subsec:HomologyM}.) Thus, ${}^\vee\Tor H_1(\Mf_3,\Z)$ as well as the isometry group are both equivalent to $\Z^n/\LatCS.$
Together, ${}^\vee\Tor H_1(\Mf_3,\Z)$ and $\Isom(\Mf_3)$ generate a noncommutative (but reducible) group that is equivalent to the operator algebra of the Wilson loops of the Chern-Simons system in (i). The subgroup ${}^\vee\Tor H_1(\Mf_3,\Z)$ corresponds to the group generated by the Wilson loops along one fixed cycle of $T^2$ (let us call it ``the $\alpha$-cycle'') and $\Isom(\Mf_3)$ corresponds to the group generated by the Wilson loops along another cycle (call it ``the $\beta$-cycle''), where $\alpha$ and $\beta$ generate $H_1(T^2,\Z)$. The situation is summarized in the following diagram:
\vskip 12pt
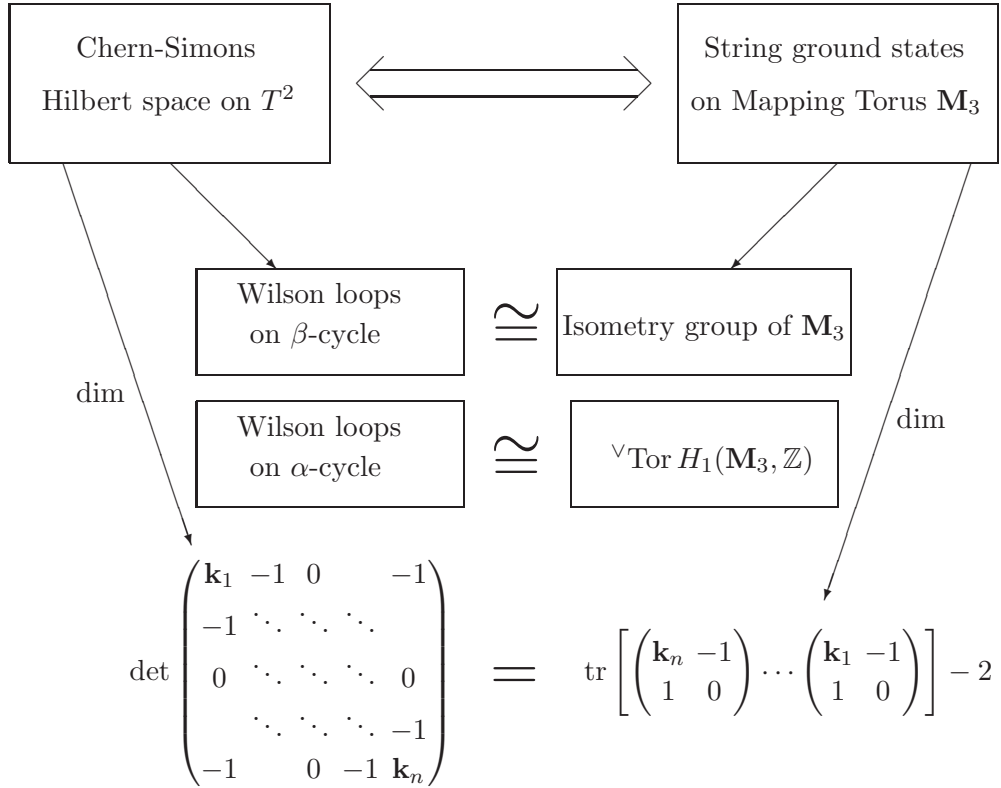
\begin{picture}(400,300)
\put(20,300){\begin{picture}(120,60)
\multiput(0,0)(0,-60){2}{\line(1,0){120}}
\multiput(0,0)(120,0){2}{\line(0,-1){60}}
\put(25,-20){Chern-Simons}
\put(12,-40){Hilbert space on $T^2$}
\put(60,-60){\vector(1,-1){40}}
\end{picture}}

\put(270,300){\begin{picture}(120,60)
\multiput(0,0)(0,-60){2}{\line(1,0){120}}
\multiput(0,0)(120,0){2}{\line(0,-1){60}}
\put(10,-20){String ground states}
\put(5,-40){on Mapping Torus $\Mf_3$}
\put(60,-60){\vector(-1,-1){40}}
\end{picture}}

\put(150,270){\begin{picture}(120,20)
\multiput(5,5)(0,-10){2}{\line(1,0){100}}
\put(0,0){\line(1,1){10}}
\put(0,0){\line(1,-1){10}}
\put(110,0){\line(-1,1){10}}
\put(110,0){\line(-1,-1){10}}
\end{picture}}

\put(90,200){\begin{picture}(100,40)
\multiput(0,0)(0,-40){2}{\line(1,0){100}}
\multiput(0,0)(100,0){2}{\line(0,-1){40}}
\put(15,-13){Wilson loops}
\put(20,-28){on $\beta$-cycle}
\end{picture}}

\put(90,150){\begin{picture}(100,40)
\multiput(0,0)(0,-40){2}{\line(1,0){100}}
\multiput(0,0)(100,0){2}{\line(0,-1){40}}
\put(15,-13){Wilson loops}
\put(20,-28){on $\alpha$-cycle}
\end{picture}}

\put(225,200){\begin{picture}(110,40)
\multiput(0,0)(0,-40){2}{\line(1,0){110}}
\multiput(0,0)(110,0){2}{\line(0,-1){40}}
\put(2,-25){Isometry group of $\Mf_3$}
\end{picture}}

\put(230,150){\begin{picture}(100,40)
\multiput(0,0)(0,-40){2}{\line(1,0){100}}
\multiput(0,0)(100,0){2}{\line(0,-1){40}}
\put(15,-25){${}^\vee\text{Tor}\,H_1(\Mf_3,\Z)$}
\end{picture}}

\put(380,240){\vector(-1,-3){55}}\put(352,140){dim}

\put(235,45){$\tr\left\lbrack
\begin{pmatrix} \lvk_n & -1 \\
1 & 0 \\
\end{pmatrix}
\cdots
\begin{pmatrix} \lvk_1 & -1 \\
1 & 0 \\
\end{pmatrix}
\right\rbrack
-2$}

\put(40,240){\vector(1,-3){48}}\put(45,150){dim}

\put(200,170){\Huge $\cong$}
\put(200,120){\Huge $\cong$}
\put(200,41){\Huge $=$}

\put(65,45){$\det\begin{pmatrix}
\lvk_1 & -1 & 0 &  & -1 \\
-1 & \ddots & \ddots & \ddots & \\
0 & \ddots & \ddots & \ddots & 0 \\
 &\ddots  & \ddots & \ddots & -1 \\
-1 &  & 0 & -1 & \lvk_n \\
\end{pmatrix}$}
\end{picture}
\vskip 12pt
We will now present a detailed account of the statements made above.
In \secref{sec:SL2Z} we construct the $\SL(2,\Z)$-twist from the QFT perspective, and in \secref{sec:CSLE} we take its low-energy limit and make connection with  $U(1)^n$ Chern-Simons theory, leading to Hilbert space (i).
In \secref{sec:mapT} we describe the dual construction of type-IIA strings on $\Mf_3$. In \secref{sec:Duality} we develop the ``dictionary'' that translates between the states and operators of (i) and (ii).
We conclude in \secref{sec:Discussion} with a brief summary of what we have found so far and a preview of the nonabelian case.

\section{The $\SL(2,\Z)$-twist}
\label{sec:SL2Z}

Our starting point is a free 3+1D $U(1)$ gauge theory with action
$$
I = \frac{1}{4\gYM^2}\int F\wedge^*F + \frac{\theta}{2\pi}\int F\wedge F,
$$
where $F=dA$ is the field strength.
As usual, we define the complex coupling constant
$$
\tau\equiv\frac{4\pi i}{\gYM^2}+\frac{\theta}{2\pi}\equiv \tau_1+i\tau_2.
$$
The $\SL(2,\Z)$ group of dualities is generated by $S$ and $T$ that act as $\tau\rightarrow -1/\tau$ and $\tau\rightarrow\tau+1$, respectively.

Let the space-time coordinates be $x_0,\dots,x_3$.
We wish to compactify direction $x_3$ on a circle (so that $0\le x_3\le 2\pi$ is a periodic coordinate), but allow $\tau$ to vary as a function of $x_3$ in such a way that 
$$
\tau(0)=\frac{\xa\tau(2\pi)+\xb}{\xc\tau(2\pi)+\xd}\,,
$$
where $\Matf\equiv\begin{pmatrix} \xa & \xb \\ \xc & \xd \\ \end{pmatrix}\in\SL(2,\Z)$ defines an electric/magnetic duality transformation.
Such a compactification contains two ingredients:
\begin{itemize}
\item
The variable coupling constant $\tau$; and
\item
The ``duality-twist'' at $x_3=0\sim 2\pi$.
\end{itemize}
We will discuss the ingredients separately, starting from the duality-twist.

The duality-twist can be described concretely in terms of an abelian Chern-Simons theory as follows.
Represent the $\SL(2,\Z)$ matrix in terms of the generators $S$ and $T$ (nonuniquely) as
\be\label{eqn:xabcd}
\begin{pmatrix} \xa & \xb \\ \xc & \xd \\ \end{pmatrix}
 = T^{k_1} S T^{k_2} S \cdots T^{k_n} S\,,
\ee
where $\lvk_1,\dots,\lvk_n$ are integers, some of which may be zero.
To understand how each of the operators $T$ and $S$ act separately, we pretend that $x_3$ is a time-direction and impose the temporal gauge condition $A_3=0$.
At any given $x_3$ the wave-function is formally $\Psi(\Atd)$, where $\Atd$ is the gauge field $1$-form on the three-dimensional space parameterized by $x_0, x_1, x_2$.
The action of the generators $S$ and $T$ on the wave-functions is then given by (see for instance \cite{Ganor:1996pe,Witten:2003ya}):
$$
S: \Psi(\Atd)\rightarrow \int e^{-\frac{i}{2\pi}\int\Atd\wedge d\Atd'}\Psi(\Atd')\cD\Atd'\,,\qquad
T:\Psi(\Atd)\rightarrow  e^{\frac{i}{4\pi}\int\Atd\wedge d\Atd}\Psi(\Atd)\,.
$$
It is now clear how to incorporate the duality twist by combining these two elements to realize the $\SL(2,\Z)$ transformation \eqref{eqn:xabcd}.
We have to add to the action a Chern-Simons term at $x_3=0$ with additional auxiliary fields $A_1,\dots,A_{n+1}$ and with action
\be\label{eqn:ICS}
I_{CS} = \tfrac{1}{4\pi}\sum_{i=1}^n k_i A_i\wedge dA_i
-\tfrac{1}{2\pi}\sum_{i=1}^n A_i\wedge dA_{i+1}\,,
\ee
and then set
$$
A_1=A\rvert_{x_3=0}\,,\qquad
A_{n+1}=A\rvert_{x_3=2\pi}\,.
$$
The second ingredient is the varying coupling constant $\tau(x_3)$.
Systems with such a varying $\tau$ are known as {\it Janus configurations} \cite{Bak:2003jk}.
They have supersymmetric extensions \cite{Clark:2005te}-\cite{Gaiotto:2008sd} where the Lagrangian of $N=4$ Super-Yang-Mills with variable $\tau$ is modified so as to preserve $8$ supercharges.
In such configurations the function $\tau(x_3)$ traces a geodesic in the hyperbolic upper-half $\tau$-plane, namely, a half-circle centered on the real axis \cite{Gaiotto:2008sd}. In this model, the surviving supersymmetry is described by parameters that vary as a function of $x_3$, so that in general the supercharges at $x_3=0$ are not equal to those at $x_3=2\pi$. This might have been a problem for us, since we need to continuously connect $x_3=0$ to $x_3=2\pi$ to form a consistent supersymmetric theory, but luckily, we also have the $\SL(2,\Z)$-twist, and as shown in \cite{Kapustin:2006pk}, in $N=4$ Super-Yang-Mills (with a fixed coupling constant $\tau$), the $\SL(2,\Z)$ duality transformations do not commute with the supercharges. Following the action of duality, the SUSY generators pick up a known phase. But as it turns out, this phase exactly matches the phase difference from $0$ to $2\pi$ in the Janus configuration. Therefore, we can combine the two separate ingredients and close the supersymmetric Janus configuration on the segment $[0,2\pi]$ with an $\SL(2,\Z)$ duality twist that connects $0$ to $2\pi$.
We describe this construction in more detail in \appref{app:Janus}.

The details of the supersymmetric action, however, will not play an important role in what follows, so we will just assume supersymmetry and proceed. Thanks to mass terms that appear in the Janus configuration (which are needed to close the SUSY algebra \cite{Gaiotto:2008sd}), at low-energy the superpartners of the gauge fields are all massive (see \appref{app:Janus}), with masses of the order of the Kaluza-Klein scale, and we can ignore them.
We will therefore proceed with a discussion of only the free $U(1)$ gauge fields.

\section{The Low-energy limit and Chern-Simons theory}
\label{sec:CSLE}
At low-energy we have to set $A_1=A_{n+1}$ in \eqref{eqn:ICS}, since the dependence of $A$ on $x_3$ is suppressed.
Then, the low-energy system is described by a 2+1D Chern-Simons action with gauge group $U(1)^n$ and action
$$
I = \tfrac{1}{4\pi}\sum_{i,j=1}^n \Kcpl_{ij}A_i\wedge dA_j\,,
$$
with coupling-constant matrix that is given by
\be\label{eqn:Kcpl}
\Kcpl\equiv
\begin{pmatrix}
\lvk_1 & -1 & 0 &  & -1 \\
-1 & \ddots & \ddots & \ddots & \\
0 & \ddots & \ddots & \ddots & 0 \\
 &\ddots  & \ddots & \ddots & -1 \\
-1 &  & 0 & -1 & \lvk_n \\
\end{pmatrix}\,.
\ee
We now make directions $x_1,x_2$ periodic, so that the theory is compactified on $T^2$, leaving only time uncompactified.
The dimension of the resulting Hilbert space of states of this compactified Chern-Simons theory is $|\det\Kcpl|$.

Next, we pick two fundamental cycles whose equivalence classes generate $H_1(T^2,\Z)$.
Let $\alpha$ be the cycle along a straight line from $(0,0)$ to $(1,0)$, and let $\beta$ be a similar cycle from $(0,0)$ to $(0,1)$, in $(x_1,x_2)$ coordinates. We define $2n$ Wilson loop operators:
$$
\Wa_j\equiv\exp\left(i\oint_\alpha A_j\right)\,,\qquad
\Wb_j\equiv\exp\left(i\oint_\beta A_j\right)\,,\qquad
j=1,\dots,n.
$$
They are unitary operators with commutation relations given by
$$
\Wa_i\Wa_j = \Wa_j\Wa_i\,,\qquad
\Wb_i\Wb_j = \Wb_j\Wb_i\,,\qquad
\Wa_i\Wb_j = e^{2\pi i(\Kcpl^{-1})_{ij}}\Wb_j\Wa_i\,.
$$
[$(\Kcpl^{-1})_{ij}$ is the $i,j$ element of the matrix $\Kcpl^{-1}$.]
In particular, for any $j=1,\dots,n$ the operator $\prod_{i=1}^n\Wa_i^{\Kcpl_{ij}}$ commutes with all $2n$ operators, and hence is a central element. In an irreducible representation, it can be set to the identity.
The $\Wa_i$'s therefore generate a finite abelian group, which we denote by $\Ga$. 
Similarly, we denote by $\Gb$ the finite abelian group generated by the $\Wb_i$'s.
Both groups are isomorphic and can be described as follows.
Let $\Lambda\subseteq\Z^n$ be the sublattice of $\Z^n$ generated by the columns of the matrix $\Kcpl$.
Then, $\Z^n/\Lambda$ is a finite abelian group and $\Ga\cong\Gb\cong\Z^n/\Lambda$, since an element of $\Z^n$ represents the powers of a monomial in the $\Wa_i$'s (or $\Wb_i$'s), and an element in $\Lambda$ corresponds to a monomial that is a central element.
We therefore map
\be\label{eqn:GaToZmodLambda}
\Ga\ni\prod_{i=1}^n\Wa_i^{N_i}\mapsto (N_1, N_2, \dots, N_n)\in\Z^n\pmod{\Lambda}\,,
\ee
and similarly
\be\label{eqn:GbToZmodLambda}
\Gb\ni\prod_{i=1}^n\Wb_i^{M_i}\mapsto (M_1, M_2, \dots, M_n)\in\Z^n\pmod{\Lambda}\,.
\ee
We denote the operator in $\Ga$ that corresponds to $\ZLv\in\Z^n/\Lambda$ by $\OpGa(\ZLv)$, and similarly we define $\OpGb(\ZLv)\in\Gb$ to be the operator in $\Gb$ that corresponds to $\ZLv.$
For $\ZLu,\ZLv\in\Z^n/\Lambda$ we define
\be\label{eqn:defCHuv}
\CHuv(\ZLu,\ZLv)\equiv e^{2\pi i\sum_{i,j}(\Kcpl^{-1})_{ij}N_i M_j}\,,\qquad(\ZLu,\ZLv\in\Z^n/\Lambda).
\ee
The definition is independent of the particular representatives $(N_1,\dots, N_n)$ or $(M_1,\dots,M_n)$ in $\Z^n/\Lambda$.
The commutation relations can then be written as
\be\label{eqn:OpGaOpGb}
\OpGa(\ZLu)\OpGb(\ZLv)=\CHuv(\ZLu,\ZLv)\OpGb(\ZLv)\OpGa(\ZLu)\,.
\ee

We recall that for any nonsingular matrix of integers $\Kcpl\in\GL(n,\Z)$, one can find matrices $P,Q\in\SL(n,\Z)$ such that
\be\label{eqn:Smith}
P\Kcpl Q = \diag(d_1, d_2, \ldots, d_n)
\ee
is a diagonal matrix, $d_1,\ldots, d_n$ are positive integers, and $d_i$ divides $d_{i+1}$ for $i=1,\dots,n-1.$
The integers $d_1,\dots, d_n$ are unique, and we have
$$
\Z^n/\Lambda\cong \Z_{d_1}\oplus\cdots\oplus\Z_{d_n}\,,
$$
where $\Z_d$ is the cyclic group of $d$ elements.
The matrix on the RHS of \eqref{eqn:Smith} is known as the {\it Smith normal form} of $\Kcpl.$
The integer $d_j$ is the greatest common divisor of all $j\times j$ minors of $\Kcpl$. For $\Kcpl$ of the form \eqref{eqn:Kcpl}, the minor that is made of rows $2,\dots,n-1$ and columns $1,\dots,n-2$ is $(-1)^{n-2}$, so it follows that $d_{n-2}=1$ and therefore also $d_1=\cdots d_{n-2}=1$. We conclude that
$$
\Ga\cong\Gb\cong\Z_{d_{n-1}}\oplus\Z_{d_n}\,.
$$

\section{Strings on a mapping torus}
\label{sec:mapT}
The system we studied in \secref{sec:SL2Z} has a dual description as the Hilbert space of ground states of strings of winding number $w=1$ (around a $1$-cycle to be defined below) on a certain type-IIA background.
We will begin by describing the background geometry and then explain in \secref{sec:Duality} why its space of ground states is isomorphic to the space of ground states of the $\SL(2,\Z)$-twisted compactification of \secref{sec:SL2Z}.

Set
\be\label{eqn:Matf}
\Matf=
\begin{pmatrix} \lvk_n & -1 \\
1 & 0 \\
\end{pmatrix}
\cdots
\begin{pmatrix} \lvk_2 & -1 \\
1 & 0 \\
\end{pmatrix}
\begin{pmatrix} \lvk_1 & -1 \\
1 & 0 \\
\end{pmatrix}
=T^{\lvk_n}S\cdots T^{\lvk_2}S T^{\lvk_1}S
\equiv\begin{pmatrix}
\xa & \xb \\ 
\xc & \xd \\
\end{pmatrix}
\in\SL(2,\Z)\,.
\ee
We will assume that $|\tr\Matf|>2$ so that $\Matf$ is a hyperbolic element of $\SL(2,\Z)$.
(The case of elliptic elements with $|\tr\Matf|<2$ was covered in \cite{Ganor:2010md}, and parabolic elements with $|\tr\Matf|=2$ are conjugate to $\pm T^\lvk$ for some $\lvk\neq 0$, and since they do not involve $S$, they are elementary.)

Let $0\le \xI\le 2\pi$ denote the coordinate on the interval $I=[0,2\pi]$ and let $(\xT_1,\xT_2)$ denote the coordinates of a point on $T^2.$ The coordinates $\xT_1$ and $\xT_2$ take values in $\R/\Z$ (so they are periodic with period $1$).
We impose the identification
\be\label{eqn:abcdEquiv}
(\xT_1, \xT_2, \xI)\sim (\xd\xT_1+\xb\xT_2, \xc\xT_1+\xa\xT_2, \xI+2\pi).
\ee
The metric is
$$
ds^2 =\xR^2 d\xI^2 + (\frac{4\pi^2\xS^2}{\tau_2})|d\xT_1 +\tau(\xI) d\xT_2|^2
$$
where $\xR$ and $\xS$ are constants, and $\tau=\tau_1 + i\tau_2$ is a function of $\xI$ (with real and imaginary parts denoted by $\tau_1$ and $\tau_2$) such that
$$
\tau(\xI-2\pi) = \frac{\xa\tau(\xI)+\xb}{\xc\tau(\xI)+\xd}\,,
$$
thus allowing for a continuous metric.

\subsection{The number of fixed points}
\label{subsec:fixedp}

We will need the number of fixed points of the $\SL(2,\Z)$ action on $T^2$, i.e., the number of solutions to:
$$
(\xT_1, \xT_2)= (\xd\xT_1+\xb\xT_2, \xc\xT_1+\xa\xT_2)\pmod{\Z^2}\,.
$$
Let $f:T^2\rightarrow T^2$ be the map given by
\be\label{eqn:f}
f:(\xT_1, \xT_2)\rightarrow (\xd\xT_1+\xb\xT_2, \xc\xT_1+\xa\xT_2)\,.
\ee
The Lefschetz fixed-point formula states that
$$
\sum_{\text{fixed point $p$}} i(p) = \sum_{j=0}^2 (-1)^j\tr(f_*\lvert H_j(T^2,\Z)) = 2-\tr\Matf = 2-\xa-\xd.
$$
The index $i(p)$ of a fixed point is given by \cite{BottTu}:
$$
i(p)=\sgn\det(\Jac(p)-\Id) = \sgn\det(\Matf-\Id)\,,
$$
where $\Jac(p)$ is the Jacobian matrix of the map $f$ at $p$. In our case, $i(p)$ is either $+1$ or $-1$ for all $p$, and therefore the number of fixed points is
$$
|2-\tr\Matf| = |\det(\Matf-\Id)|=|2-\xa-\xd|\,.
$$

\subsection{Isometries}
\label{subsec:IsometryM}
Let $\Iv_1, \Iv_2\in\R/\Z$ be constants  and consider the map
\be\label{eqn:IsometryIv}
(\xT_1, \xT_2, \xI)\mapsto (\xT_1+\Iv_1, \xT_2+\Iv_2, \xI).
\ee
It defines an isometry of $\Mf_3$ if
\be\label{eqn:abcdvv}
\begin{pmatrix}
\xa & \xc \\ 
\xb & \xd \\
\end{pmatrix}
\begin{pmatrix}
\Iv_2 \\ 
\Iv_1 \\
\end{pmatrix}
\equiv
\begin{pmatrix}
\Iv_2\\ 
\Iv_1 \\
\end{pmatrix}
\qquad\pmod\Z
\,.
\ee
Set
\be\label{eqn:defWmId}
\WmId\equiv \Matf^T-\Id=\begin{pmatrix}
\xa-1 & \xc \\ 
\xb & \xd-1 \\
\end{pmatrix}\,,
\qquad
\IvVec\equiv\begin{pmatrix}
\Iv_2\\ 
\Iv_1 \\
\end{pmatrix}\,.
\ee
Then, the isometries are given by
$
\IvVec=\WmId^{-1}
\begin{pmatrix}
n_2\\ 
n_1 \\
\end{pmatrix}
$
for some $n_1, n_2\in\Z$.
The set of vectors $\IvVec$ that give rise to isometries therefore live on a lattice $\widetilde{\Lambda}$ generated by the columns of $\WmId^{-1}$. Since $\WmId\in\GL(2,\Z)$ we have $\Z^2\subseteq\widetilde{\Lambda}$, and since the isometries that correspond to $\IvVec\in\Z^2$ are trivial, the group of isometries of type \eqref{eqn:IsometryIv} is isomorphic to $\widetilde{\Lambda}/\Z^2$. Changing basis to $\IuVec\equiv\WmId\IvVec$, we can replace $\IvVec\in\widetilde{\Lambda}/\Z^2$ with $\IuVec\in\Z^2/\Lambda'$, where $\Lambda'\subseteq\Z^2$ is the sublattice generated by the columns of $\WmId$, and the group $\Giso$ of isometries of type \eqref{eqn:IsometryIv} is therefore
\be\label{eqn:Giso}
\Giso
\cong\widetilde{\Lambda}/\Z^2
\cong\Z^2/\Lambda'\,.
\ee
Its order is 
\be\label{eqn:GisoOrder}
|\Giso|=|\det\WmId|=|2-\xa-\xd|.
\ee

\subsection{Homology quantum numbers}
\label{subsec:HomologyM}

To proceed we also need the homology group $H_1(\Mf_3,\Z)$.
Let $\gamma$ be the cycle defined by a straight line from $(0,0,0)$ to $(0,0,2\pi)$, in terms of $(\xT_1,\xT_2,\xI)$ coordinates. Let $\alpha'$ be the cycle from $(0,0,0)$ to $(1,0,0)$ and let $\beta'$ be the cycle from $(0,0,0)$ to $(0,1,0)$.
The homology group $H_1(\Mf_3,\Z)$ is generated by the equivalence classes $[\alpha']$, $[\beta']$ and $[\gamma]$, subject to the relations
\be\label{eqn:H1relations}
[\alpha']  =\xd[\alpha']+\xc[\beta']\,,\qquad [\beta']=\xb[\alpha']+\xa[\beta'].
\ee
Now suppose that $(c_1\, c_2)$ is a linear combination of the columns of $\WmId$ [defined in \eqref{eqn:defWmId}] with integer coefficients. Then the relations \eqref{eqn:H1relations} imply that $c_1[\alpha']+c_2[\beta']$ is zero in $H_1(\Mf_3,\Z)$. With $\Lambda'\subset\Z^2$ being the sublattice generated by the columns of $\WmId$, as defined in  \secref{subsec:IsometryM}, it follows that 
\be\label{eqn:H1}
H_1(\Mf_3,\Z)\cong\Z\oplus(\Z^2/\Lambda'),
\ee
where the $\Z$ factor is generated by $[\gamma]$ and $(\Z^2/\Lambda')$ is generated by $[\alpha']$ and $[\beta']$.
In particular, the torsion part is
\be\label{eqn:TorH1}
\Tor H_1(\Mf_3,\Z)\cong\Z^2/\Lambda'\,.
\ee

Denote the Smith normal form [see \eqref{eqn:Smith}] of the matrix $\WmId$ by
$\begin{pmatrix} d_1' & \\ & d_2' \\ \end{pmatrix}.$
We prove in \appref{app:DetSmithProof} that $d_{n-1}=d_1'$ and $d_n=d_2'$, where $d_{n-1}$ and $d_n$ were defined in \eqref{eqn:Smith}. Thus, combining \eqref{eqn:Giso} and \eqref{eqn:H1} we have
$$
\Z^2/\Lambda'\cong
\Giso\cong\Tor H_1(\Mf_3,\Z)\cong
\Z_{d_{n-1}}\oplus\Z_{d_n}\,.
$$
The physical meaning of these results will become clear soon.

\subsection{The Hilbert space of states}
\label{subsec:StringStates}

As we have seen in \secref{subsec:IsometryM}, the Hilbert space of string ground states has a basis of states of the form
$\ket{\IvVec'}$ with $\IvVec'\in\widetilde{\Lambda}/\Z^2$. In this state, the string is at $(\xT_1, \xT_2)$ coordinates given by $\IvVec'$.
According to \eqref{eqn:Giso},  an element $\IvVec\in\widetilde{\Lambda}/\Z^2$ defines an isometry, which we denote by $\Iso(\IvVec)$, that acts as
$$
\Iso(\IvVec)\ket{\IvVec'} = \ket{\IvVec+\IvVec'}\,,\qquad
\IvVec, \IvVec'\in\widetilde{\Lambda}/\Z^2\,.
$$
Given the string state $\ket{\IvVec'}$, we can ask what is the element in $H_1(\Mf_3,\Z)$ that represents the corresponding $1$-cycle. The answer is $[\gamma]+N'_1[\alpha']+N'_2[\beta']$, where the torsion part $N'_1[\alpha']+N'_2[\beta']$ is mapped under \eqref{eqn:TorH1} to $\IvVec'$.
To see this, note that for $0\le t\le 1$, the loops $C_t$ in $\Mf_3$ that are given by
$$
 \left\{
\begin{array}{ll}
(4\pi s, t\Iv_1', t\Iv_2') & \text{for $0\le s\le \tfrac{1}{2}$} \\
(2\pi, t\Iv_1' + (2s-1)t [(\xd-1)\Iv_1' + \xb\Iv_2'], t\Iv_2' + (2s-1)t [\xc\Iv_1' + (\xa-1)\Iv_2'])
& \text{for $\tfrac{1}{2}\le s\le 1$} \\
\end{array}\right.
$$
[which go along direction $\xI$ at a constant $(\xT_1,\xT_2)$ given by $t\IvVec'$, and then connect $t\IvVec'$ to its $\SL(2,\Z)$ image $t\Matf\IvVec'$] are homotopic to the loop corresponding to string state $\ket{0}$.
Setting $t=1$ we find that $C_1$ breaks into two closed loops, one corresponding to string state $\ket{\IvVec'}$, and the other is a closed loop in the $T^2$ fiber above $\xI=0$, which corresponds to the homology element
$$
((\xd-1)\Iv_1' + \xb\Iv_2')[\alpha']+(\xc\Iv_1' + (\xa-1)\Iv_2')[\beta']\,,
$$
and this is precisely the element corresponding to $\WmId\IvVec'\in\Z^2/\Lambda'\cong\Tor H_1(\Mf_3,\Z)$, as defined in \secref{subsec:HomologyM}.

We now wish to use the torsion part of the homology to define a unitary operator $\HChg(\IuChg)$ for every $\IuChg\in\Z^2/\Lambda'$. This operator will measure a component of the charge associated with the homology class of the string. For this purpose we need to construct the Pontryagin dual group ${}^\vee\Tor H_1(\Mf_3,\Z)$, which is defined as the group of characters of $\Tor H_1(\Mf_3,\Z)$ (i.e., homomorphisms from $\Tor H_1(\Mf_3,\Z)$ to $\R/\Z$). The dual group is isomorphic to $\Z^2/\Lambda'$, but not canonically. In our construction $\IuChg$ is naturally an element of the dual group and not the group itself. We define $\HChg(\IuChg)$ as follows.
For
$$
\IuChg=(M_1', M_2')\in\Z^2/\Lambda'\,,\qquad
\IvVec=(N_1', N_2')\in\Z^2/\Lambda'\,,
$$
we define the phase
\be\label{eqn:defPhuv}
\Phuv(\IuChg,\IvVec)\equiv e^{2\pi i (\WmId^{-1})_{ij} N_i' M_j'}\,,\qquad
\IuChg\in\Z^2/\Lambda'\,,\quad
\IvVec\in\Z^2/\Lambda'\,.
\ee
This definition is independent of the representatives $(N_1', N_2')$ and $(M_1', M_2')$ of $\IvVec$ and $\IuChg$, and it corresponds to the character of $\Tor H_1(\Mf_3,\Z)$ associated with $\IuChg$.
We then define the operator $\HChg(\IuChg)$ to be diagonal in the basis $\ket{\IvVec}$ and act as:
$$
\HChg(\IuChg)\ket{\IvVec} = \Phuv(\IuChg,\IvVec)\ket{\IvVec}\,,\qquad
\IuChg\in\Z^2/\Lambda'\,,\quad
\IvVec\in\Z^2/\Lambda'\,.
$$
{}From the discussion above about the homology of the string state, and from the linearity of the phase of $\Phuv(\IuChg,\IvVec)$ in $\IuChg$ and $\IvVec$, it follows that
\be\label{eqn:HChgIso}
\HChg(\IuChg)\Iso(\IvVec)
=\Phuv(\IuChg,\IvVec)\Iso(\IvVec)\HChg(\IuChg)\,.
\ee

\section{Duality between strings on $\Mf_3$ and compactified $\SL(2,\Z)$-twisted $U(1)$ gauge theory}
\label{sec:Duality}

We can now connect the string theory model of \secref{sec:mapT} with the field theory model of \secref{sec:CSLE}.
We claim that the Hilbert space of ground states of a compactification of a $U(1)$ gauge theory on $S^1$ with an $\SL(2,\Z)$ twist and string ground states on $\Mf_3$ are dual. This is demonstrated along the same lines as in \cite{Ganor:2010md}.
We realize the (supersymmetric extension of the) $U(1)$ gauge theory on a D$3$-brane along directions $x_1, x_2, x_3.$ We compactify direction $x_3$ on a circle with a Janus-like configuration and $\SL(2,\Z)$-twisted boundary conditions. 
We assume that the Janus configuration can be lifted to type-IIB, perhaps with additional fluxes, but we will not worry about the details of the lift. We then compactify $(x_1,x_2)$ on $T^2$ and perform T-duality on direction $1$, followed by a lift from type-IIA to M-theory (producing a new circle along direction $10$), followed by reduction to type-IIA along direction $2$. This combined U-duality transformation transforms the $\SL(2,\Z)$-twist to the geometrical transformation \eqref{eqn:abcdEquiv}.
It also transforms some of the charges of the type-IIB system to the following charges of the type-IIA system:
\be\label{eqn:ChargeMap}
D3_{123}\rightarrow F1_3\,,\quad
F1_1\rightarrow P_1\,,\quad
F1_2\rightarrow F1_{10}\,,\quad
D1_1\rightarrow F1_1\,,\quad
D1_2\rightarrow P_{10}\,.
\ee
where $P_j$ is Kaluza-Klein momentum along direction $j$, D$p_{j_1\ldots j_r}$ is a D$p$-brane wrapped along directions $j_1,\ldots,j_r$, and F$1_j$ is a fundamental string along direction $j$.

Now suppose we take the limit that all directions of $\Mf_3$ are large.
The dual geometry has a Hilbert space of ground states which corresponds to classical configurations of strings of minimal length that wind once around the $x_3$ circle. This means that the projection of their $H_1(\Mf_3,\Z)$ homology class on the $\Z$ factor of \eqref{eqn:H1} is required to be the generator $[\gamma]$. The torsion part of their homology is unrestricted.
The string configurations of minimal length must have constant $(x_1, x_2)$ which in particular means that $(x_1,x_2)$ is invariant under the $\SL(2,\Z)$ twist, i.e.,
$$
\begin{pmatrix}
\xa & \xc \\ 
\xb & \xd \\
\end{pmatrix}
\begin{pmatrix}
x_2 \\ 
x_1 \\
\end{pmatrix}
\equiv
\begin{pmatrix}
x_2\\ 
x_1 \\
\end{pmatrix}
\pmod\Z
\,.
$$
But this is precisely the same equation as \eqref{eqn:abcdvv}, and indeed when the isometry that corresponds to a vector $\IvVec\in\widetilde{\Lambda}/\Z^2$ acts on the solution with $(x_1, x_2)=(0,0)$ it converts it to the solution with $(x_1,x_2)=(\Iv_1,\Iv_2).$ The dimension of the Hilbert space of ground states of the type-IIA system is therefore the order of $\Giso$, which is given by \eqref{eqn:GisoOrder}.
This is also the number of fixed points of the $\Matf$ action on $T^2$, as we have seen in \secref{subsec:fixedp}.
Since the number of ground states of the Chern-Simons theory is $|\det\Kcpl|$, we conclude from the duality of the Chern-Simons theory and string theory that
$$
|\det\Kcpl| =|\Giso| = |2-\xa-\xb|\,.
$$
This is the physical explanation we are giving to \eqref{eqn:MolinariSpecial}.

\subsection{Isomorphism of operator algebras}

Going one step beyond the equality of dimensions of the Hilbert spaces, we would like to match the operator algebras of the string and field theory systems.
Starting with the field theory side, realized on a D$3$-brane in type-IIB, consider a process whereby a fundamental string that winds once around the $\beta$-cycle of $T^2$ is absorbed by the D$3$-brane at some time $t$.
This process is described in the field theory by inserting a Wilson loop operator $\Wb_1$ at time $t$ into the matrix element that calculates the amplitude.
On the type-IIA string side, the charge $F1_2$ that was absorbed is mapped by \eqref{eqn:ChargeMap} to winding number along the $\alpha'$ cycle (denoted by $F1_{10}$).
The operator that correpsonds to $\Wb_1$ on the string side must therefore increase the homology class of the string state by $[\alpha']$. Since the state $\ket{\IvVec}$, for $\IvVec=(N_1', N_2')$, has homology class $[\gamma]+N_1'[\alpha']+N_2'[\beta']$, it follows that the isometry operator $\Iso(\IvVec')$ with $\IvVec'=(1,0)$ does what we want. We therefore propose to identify
$$
\Wb_1\rightarrow\Iso(\IvVec')\,,\qquad\text{for $\IvVec'=(1,0).$}
$$
By extension, we propose to map the abelian subgroup $\Gb$ generated by the Wilson loops $\Wb_1,\dots,\Wb_n$ with the isometry group generated by $\Iso(\IvVec')$ for $\IvVec'\in\Z^2/\Lambda'$.

Next, on the type-IIB side,  consider a process whereby a fundamental string that winds once around the $\alpha$-cycle of $T^2$ is absorbed by the D$3$-brane.
This process is described in the field theory by inserting a Wilson loop operator $\Wa_1$ into the matrix element that calculates the amplitude.
On the type-IIA string side, the charge $F1_1$ that was absorbed is mapped by \eqref{eqn:ChargeMap} to momentum along the $\beta'$ cycle (denoted by $P1_1$). The operator that correpsonds to $\Wa_1$ on the string side must therefore increase the momentum along the $[\alpha']$ cycle by one unit. We claim that this operator is $\HChg(\IuChg)$ for $\IuChg=(1,0).$
To see this we note that, by definition of ``momentum'', an operator $\Op$ that increases the momentum by $M_1'$ units along the $[\alpha']$ cycle and $M_2'$ units along the $[\beta']$ cycle must have the following commutation relations with the translational isometries $\Iso(\IvVec')$:
$$
\Iso(\IvVec')^{-1}\Op\Iso(\IvVec')=\Phuv(\IuChg,\IvVec')\Op\,,\qquad
\IuChg = (M_1',M_2')\in\Z^2/\Lambda'.
$$
But given \eqref{eqn:HChgIso}, this means that up to an unimportant central element, we can identify $\Op=\HChg(\IuChg)$, as claimed. 
So, we have
$$
\Wa_1\rightarrow\Iso(\IuChg)\,,\qquad\text{for $\IuChg=(1,0),$}
$$
and by extension, we propose to map the abelian subgroup $\Ga$ generated by the Wilson loops $\Wa_1,\dots,\Wa_n$ with the subgroup generated by $\HChg(\IuChg)$ for $\IuChg\in\Z^2/\Lambda'$.

In particular, $\Ga\cong\Gb\cong\Z^n/\Lambda$ implies that $(\Z^2/\Lambda')\cong(\Z^n/\Lambda)$.
This is equivalent to requiring that the Smith normal form of $\WmId$ is
$$
P'\WmId Q' = \diag(d_{n-1}, d_n)
$$
where $d_{n-1}$ and $d_n$ are the same last two entries in the Smith normal form of $\Kcpl$.
We provide an elementary proof of this fact in \appref{app:DetSmithProof}.

Since the Smith normal forms of $\WmId$ and $\Kcpl$ are equal, the abelian groups $\Z^n/\Lambda$ and $\Z^2/\Lambda'$ are equivalent, and it is also not hard to see that under this equivalence $\CHuv$ that was defined in \eqref{eqn:defCHuv} is mapped to $\Phuv$ defined in \eqref{eqn:defPhuv}.
We have the mapping
$$
\OpGa(\ZLv)\rightarrow\Iso(\IvVec')\,,\qquad
\ZLv\in\Z^n/\Lambda\,,\qquad
\IvVec'\in\Z^2/\Lambda'
$$
and
$$
\OpGb(\ZLu)\rightarrow\HChg(\IuChg)\,,\qquad
\ZLu\in\Z^n/\Lambda\,,\qquad
\IuChg\in\Z^2/\Lambda'.
$$
The commutation relations \eqref{eqn:OpGaOpGb} are then mapped to \eqref{eqn:HChgIso}.

\section{Discussion}
\label{sec:Discussion}

We have argued that a duality between $U(1)^n$ Chern-Simons theory on $T^2$ with coupling constant matrix \eqref{eqn:Kcpl} and string configurations on a mapping torus provide a geometrical realization to the algebra of Wilson loop operators in the Chern-Simons theory.
Wilson loop operators along one cycle of $T^2$ correspond to isometries that act as translations along the fiber of the mapping torus, while Wilson loop operators along the other cycle correspond to discrete charges that can be constructed from the homology class of the string.

These ideas have an obvious extension to the case of $U(N)$ gauge group with $N>1$, where $\SL(2,\Z)$-duality is poorly understood.
The techniques presented in this paper can be extended to construct the algebra of Wilson loop operators. The Hilbert space on the string theory side is constructed from string configurations on a mapping torus whose $H_1(\Mf_3,\Z)$ class maps to $N$ under the projection map $\Mf_3\rightarrow S^1$. In other words, the homology class projects to $N[\gamma]$ when the torsion part is ignored. Such configurations could be either a single-particle string state wound $N$ times, or a multi-particle string state. A string state with $r$ strings with winding numbers $N_1,\dots, N_r$ is described by a partition $N=N_1 + \cdots N_r$, and the $j^{th}$ single-particle string state is described by an unordered set of $N_j$ points on $T^2$ that is invariant, as a set, under the action of $f$ in \eqref{eqn:f}. The counterparts of the Wilson loops on the string theory side can then be constructed from operations on these sets. A more complete account of the nonabelian case will be reported elsewhere  \cite{toAppear}.

It is interesting to note that some similar ingredients to the ones that appear in this work also appeared in \cite{Gadde:2013sca} in the study of vacua of compactifications of the free $(2,0)$ theory on Lens spaces. More specifically, a Chern-Simons theory with a tridiagonal coupling constant matrix and the torsion part of the first homology group played a role there as well. It would be interesting to further explore the connection between these two problems.

\section*{Acknowledgements}
We are grateful to T.~Dimofte, S.~Gukov, H.-S.~Tan, and R.~Thorngren for very helpful discussions and comments.
This research was supported by the Berkeley Center of Theoretical Physics. The research of NM was also supported in part by the U.S. National Science Foundation under grant No. PHY-10-02399. The research of NTC was supported by the National Science Foundation Graduate Research Fellowship Program under Grant No. DGE-11-06400.

\begin{appendix}

\section{A proof of the determinant identity and the Smith normal form of the coupling constant matrix}
\label{app:DetSmithProof}

Molinari gave an elegant proof \cite{Molinari:2007} to a generalization of \eqref{eqn:MolinariSpecial} using only polynomial analysis.
Here we present an alternative basic linear-algebra proof for \eqref{eqn:MolinariSpecial}.
At the same time we also demonstrate that the Smith normal form of the coupling constant matrix $\Kcpl$ defined in \eqref{eqn:Kcpl},
$$
\Kcpl=
\begin{pmatrix}
\lvk_1 & -1 & 0 & \ddots & -1 \\
-1 & \ddots & \ddots & \ddots & \ddots\\
0 & \ddots & \ddots & \ddots & 0 \\
\ddots &\ddots  & \ddots & \ddots & -1 \\
-1 & \ddots & 0 & -1 & \lvk_n \\
\end{pmatrix}\,,
$$
is identical to the Smith normal form of 
$$
\WmId = \Matf-\Id=\begin{pmatrix}
\xa-1 & \xb \\ 
\xc & \xd-1 \\
\end{pmatrix}\,,
$$
where $\Matf$ was defined in \eqref{eqn:Matf}.

We begin by moving the first row of $\Kcpl$ to the end, to get $\Kcpl'_1$.
We have 
$$
\det\Kcpl = (-1)^n\det\Kcpl'_1
$$
but both $\Kcpl$ and $\Kcpl'_1$ have the same Smith normal form.
For clarity, we will present explicit matrices for the $n=5$ case. We get:
$$
\Kcpl'_1\equiv
\begin{pmatrix}
-1 & \lvk_2 & -1 & 0 & 0 \\
0 & -1 & \lvk_3 & -1 & 0 \\
0 & 0 & -1  & \lvk_4 & -1 \\
-1 & 0 & 0 & -1 & \lvk_5 \\
\lvk_1 & -1 & 0 & 0 & -1 \\
\end{pmatrix}\,,
$$

We will now show how to successively define a series of matrices
$$
\Kcpl_2',\ldots,
\Kcpl_{n-1}' = \begin{pmatrix}
-1 & & & & \\
& \ddots & & & \\
& & -1 & & \\
& & & \xa-1 & \xb \\ 
& & & \xc & \xd-1 \\
\end{pmatrix}\,,
$$
related to each other by row and column operations that preserve the Smith normal form.
At each step, we need to keep track of a $2\times 2$ block of $\Kcpl_m'$ formed from the elements on the $(n-1)^{th}$ and $n^{th}$ rows and the $m^{th}$ and $(m+1)^{st}$ columns.
$$
\WmId'_m\equiv
\begin{pmatrix}
[\Kcpl_m']_{(n-1)\,m} & [\Kcpl_m']_{(n-1)\,(m+1)}   \\
[\Kcpl_m']_{n\,m} & [\Kcpl_m']_{n\,(m+1)} \\
\end{pmatrix}
$$
At the outset we have
$$
\WmId'_1\equiv
\begin{pmatrix}
[\Kcpl_1']_{(n-1)\,1} & [\Kcpl_1']_{(n-1)\,2}&  \\
[\Kcpl_1']_{n\,1} & [\Kcpl_1']_{n\,2} \\
\end{pmatrix}
=
\begin{pmatrix}
-1 & 0  \\
\lvk_1  & -1 \\
\end{pmatrix}\,.
$$
As will soon be clear from the construction, the matrix $\Kcpl_{m}'$ has the following block form:
\be\label{eqn:Kcplm}
\Kcpl'_{m} = \begin{pmatrix}
-\Id_{m-1} &  &    &              &  &  & \\
         &-1 &\lvk_{m+1} & -1 & * &* & *\\
        &   &  -1            & \lvk_{m+2} & * & * & *\\
        &      &               &  & X_{n-m-4} & * & * \\
        &[\WmId'_m]_{11} & [\WmId'_m]_{12} & & & -1 &  \lvk_n \\
        &[\WmId'_m]_{21} & [\WmId'_m]_{22} & & &   & -1\\
\end{pmatrix}\,,
\ee
where $\Id_{m-1}$ is the $(m-1)\times (m-1)$ identity matrix,
$*$ represents a block of possibly nonzero elements, $X_{n-m-4}$ represents a nonzero $(n-m-4)\times(n-m-4)$ matrix
and empty positions are zero.
To get $\Kcpl_{m+1}'$ from $\Kcpl_m'$ we perform the following row and column operations on $\Kcpl_m'$:
\begin{itemize}
\item
Add $[\WmId'_m]_{11}$ times the $m^{th}$ row to the $(n-1)^{st}$ row;
\item
Add $[\WmId'_m]_{21}$ times the $m^{st}$ row to the $n^{th}$ row;
\item
For $j=m+1,\dots,n$, add $[\Kcpl_m']_{m j}$ times the $m^{th}$ column to the $j^{th}$ column.
\end{itemize}
It is not hard to see that these operations produce a matrix that fits the general form \eqref{eqn:Kcplm} with $m\rightarrow m+1$.
Tracking how the bottom two rows transform, we find that for $m<n-2$,
$$
\WmId'_{m+1}=
\begin{pmatrix}
[\WmId'_{m+1}]_{11} & [\WmId'_{m+1}]_{12} \\
[\WmId'_{m+1}]_{21} & [\WmId'_{m+1}]_{22} \\
\end{pmatrix}
=
\begin{pmatrix}
 [\WmId'_m]_{12} +\lvk_{m+1} [\WmId'_m]_{11} &\quad -[\WmId'_m]_{11}\\
 [\WmId'_m]_{22} + \lvk_{m+1} [\WmId'_m]_{21} &\quad -[\WmId'_m]_{21}\\
\end{pmatrix}
=
\WmId'_m
\begin{pmatrix}
\lvk_{m+1} & 1 \\
-1 & 0\\
\end{pmatrix}\,.
$$
Since, by definition,
$
\WmId_1' = 
\begin{pmatrix}
-1 & 0 \\
\lvk_1 & -1\\
\end{pmatrix}\,
$,
it follows that
$$
\WmId_{n-2}' = 
\begin{pmatrix}
-1 & 0 \\
\lvk_1 & -1\\
\end{pmatrix}
\begin{pmatrix}
\lvk_2 & 1 \\
-1 & 0\\
\end{pmatrix}
\cdots
\begin{pmatrix}
\lvk_{n-2} & 1 \\
-1 & 0\\
\end{pmatrix}\,.
$$
It can then be easily checked that the last two steps yield:
$$
\WmId_n' = 
\WmId_{n-2}'
\begin{pmatrix}
\lvk_{n-1} & 1 \\
-1 & 0\\
\end{pmatrix}
\begin{pmatrix}
\lvk_n & 1 \\
-1 & 0\\
\end{pmatrix}
-
\begin{pmatrix}
1 & 0 \\
0 & 1 \\
\end{pmatrix}
\,.
$$

\section{Compatibility of the supersymmetric Janus configuration and the duality twist}
\label{app:Janus}

In this section we describe the details of the supersymmetric Lagrangian.
As explained in \secref{sec:SL2Z}, the system is composed of two ingredients:
(i) the supersymmetric Janus configuration; and (ii) an $\SL(2,\Z)$ duality twist.
We will now review the details of both ingredients and demonstrate that their combination preserves supersymmetry.

\subsection{Supersymmetric Janus}
Extending the work of \cite{Bak:2003jk}-\cite{D'Hoker:2006uv}, Gaiotto and Witten \cite{Gaiotto:2008sd} have constructed a supersymmetric deformation of \SUSY{4} Super-Yang-Mills theory with a complex coupling constant $\tau$ that varies along one direction, which we denote by $x_3$.
We will now review this construction, using the same notation as in \cite{Gaiotto:2008sd}.
First, the real and imaginary parts of the coupling constant are defined as
\be\label{eqn:GWtau}
\tau = \frac{\theta}{2\pi}+\frac{2\pi i}{\GWe^2}\,,\qquad
\ee
It is taken to vary along a semi-circle on the upper half $\tau$-plane, centered on the real axis:
\be\label{eqn:semiCircle}
\tau = \GWa + 4\pi\GWD e^{2 i\GWpsi}\,,\qquad
\ee
where $\GWpsi(x_3)$ is an arbitrary function.

The action is defined as
$$
I = I_{\text{N=4}} +I' + I'' + I'''
$$
where $I_{\text{N=4}}$ is the standard \SUSY{4} action, modified only by making $\tau$ a function of $x_3$, and $I'$, $I''$, and $I'''$ are correction terms listed below.
We will list the actions for a general gauge group, as derived by Gaiotto and Witten, although the application in this paper is for a $U(1)$ gauge group, and so several terms drop out.
The bosonic fields are: a gauge field $A_\mu$ ($\mu=0,1,2,3$), $3$ adjoint-valued scalar fields $X^a$ ($a=1,2,3$) and $3$ adjoint-valued scalar fields $Y^p$ ($p=1,2,3$). In the $U(1)$ case, $X^a$ and $Y^p$ are real scalar fields. In the type-IIB realization on  D$3$-branes, the D$3$-brane is in directions $0,1,2,3$, $X^a$ corresponds to fluctuations in directions $4,5,6$, and $Y^p$ corresponds to directions $7,8,9.$
The fermionic fields are encoded in a $16$-dimensional Majorana-Weyl spinor $\Psi$ on which even products of the 9+1D Dirac matrices $\Gamma_0,\dots,\Gamma_9$ act. Products of pairs from the list $\Gamma_0,\dots,\Gamma_3$ correspond to generators of the Lorentz group $SO(1,3)$, while products of pairs from the list $\Gamma_4,\Gamma_5,\Gamma_6$ correspond to generators of the R-symmetry subgroup $SO(3)_X$ acting on $X^1, X^2, X^3$, and products of pairs from the list $\Gamma_7,\Gamma_8,\Gamma_9$ correspond to generators of the R-symmetry subgroup $SO(3)_Y$ acting on $Y^1, Y^2, Y^3$.
We have the identity $\Gamma_{0123456789} = 1.$

The additional terms are
\bear
I' &=&
\frac{i}{\GWe^2}\int d^4x\Tr\bPsi (\GWalpha\Gamma_{012}+\GWbeta\Gamma_{456}+\GWgamma\Gamma_{789})\Psi
\,,\nn\\
I'' &=&
\frac{1}{\GWe^2}\int d^4x\Tr\left(
\GWu\epsilon^{\mu\nu\lambda}(A_\mu\partial_\nu A_\lambda +\tfrac{2}{3}A_\mu A_\nu A_\lambda)
+\tfrac{\GWv}{3}\epsilon^{abc} X_a \lbrack X_b, X_c\rbrack
+\tfrac{\GWw}{3}\epsilon^{pqr} Y_p \lbrack Y_q, Y_r\rbrack
\right)
\,,\nn\\
I''' &=&
\frac{1}{2\GWe^2}\int d^4x\Tr\left(
\GWr X_a X^a + \GWtdr Y_p Y^p
\right)
\,,\nn
\eear
where
\be\label{eqn:GWualpha}
-\tfrac{1}{4}\GWu=\GWalpha = -\tfrac{1}{2}\GWpsi'\,,
\qquad
-\tfrac{1}{4}\GWv=\GWbeta =-\frac{\GWpsi'}{2\cos\GWpsi}\,,\qquad
-\tfrac{1}{4}\GWw=\GWgamma = \frac{\GWpsi'}{2\sin\GWpsi}\,,
\ee

\be\label{eqn:GwrGwtdr}
\GWr = 2(\GWpsi'\tan\GWpsi)' + 2(\GWpsi')^2\,,\qquad
\GWtdr = -2(\GWpsi'\cot\GWpsi)' + 2(\GWpsi')^2\,.
\ee

As we are working with a $U(1)$ gauge group, we will not need the cubic terms in $I''$. They are nevertheless listed here for reference, and they will become relevant for extensions to a nonabelian gauge group.

To describe the preserved supersymmetry we follow Gaiotto-Witten and work in a spinor representation where
$$
\Gamma_{0123}=-\Gamma_{456789} 
= \begin{pmatrix} 
0 & -\Id \\
\Id & 0\\
\end{pmatrix}
\,,\quad
\Gamma_{3456}= \begin{pmatrix} 
0 & \Id \\
\Id & 0\\
\end{pmatrix}
\,,\quad
\Gamma_{3789}=
\begin{pmatrix} 
\Id & 0 \\
0 & -\Id\\
\end{pmatrix}
\,,
$$
where $\Id$ is an $8\times 8$ identity matrix.
The surviving supersymmetries are those parameterized by a $16$-component $\sParamEpsilon_{16}$ which takes the form
\be\label{eqn:sParampsi}
\sParamEpsilon_{16} = \begin{pmatrix}
\cos(\frac{\GWpsi}{2})\sParamEpsilon_8
\\
\sin(\frac{\GWpsi}{2})\sParamEpsilon_8
\\
\end{pmatrix}
\,,
\ee
where $\sParamEpsilon_8$ is an arbitrary constant $8$-component spinor.

\subsection{Introducing an $\SL(2,\Z)$-twist}
\begin{figure}[t]
\begin{picture}(400,100)
\put(10,10){\vector(1,0){380}}\put(392,5){$\tau_1$}
\put(200,10){\vector(0,1){80}}\put(196,95){$\tau_2$}
\put(230,10){\begin{picture}(100,100)
\put(60,0){\circle*{1}}
\put(59,10){\circle*{1}}
\put(56,21){\circle*{1}}
\put(52,30){\circle*{1}}
\put(-52,30){\circle*{1}}
\put(-56,21){\circle*{1}}
\put(-59,10){\circle*{1}}
\put(-60,0){\circle*{1}}

\thicklines
\qbezier(46,39)(38,48)(30,52)
\qbezier(30,52)(21,57)(10,59)
\qbezier(10,59)(0,61)(-10,59)
\qbezier(-10,59)(-21,57)(-30,52)
\qbezier(-30,52)(-38,48)(-46,39)

\put(0,60){\vector(-1,0){0}}
\put(-46,39){\circle*{3}}
\put(46,39){\circle*{3}}

\thinlines
\put(0,0){\line(0,-1){5}}\put(-3,-12){$\GWa$}
\put(60,0){\line(0,-1){5}}\put(57,-12){$\GWa+4\pi\GWD$}
\put(-60,0){\line(0,-1){5}}\put(-63,-12){$\GWa-4\pi\GWD$}

\put(-75,36){$\tau(2\pi)$}
\put(48,36){$\tau(0)$}
\qbezier(-43,39)(13,49)(0,39)
\qbezier(43,39)(-13,29)(0,39)
\put(43,39){\vector(3,1){0}}

\put(-10,25){$\tau\rightarrow\frac{\xa\tau+\xb}{\xc\tau+\xd}$}
\end{picture}}

\end{picture}
\caption{
In the Janus configuration the coupling constant $\tau$ traces a portion of a semi-circle of radius $4\pi\GWD$ in the upper-half plane, whose center $\GWa$ is on the real axis.
We augment it with an $\SL(2,\Z)$ duality twist that glues $x_3=2\pi$ to $x_3=0$.
}
\label{fig:JanusAndSL2Z}
\end{figure}
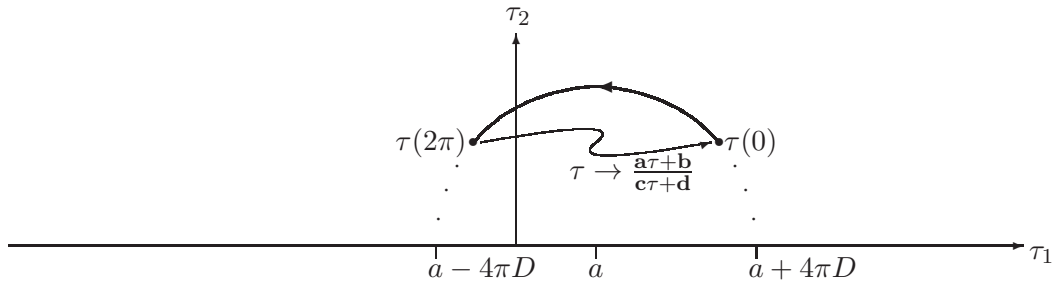

Here $\GWpsi$ is a function of $x_3$ such that $\tau(x_3)$ traces a geodesic on $\tau$-plane with metric $|d\tau|^2/\tau_2^2$.
We pick the parameters $a$ and $D$ so that 
the semi-circle \eqref{eqn:semiCircle} will be invariant under
$$
\tau\rightarrow \frac{\xa\tau+\xb}{\xc\tau+\xd}\,.
$$
This amounts to solving the two equations
$$
(\GWa-4\pi\GWD) = \frac{\xa(\GWa-4\pi\GWD)+\xb}{\xc(\GWa-4\pi\GWD)+\xd}\,,\qquad
(\GWa+4\pi\GWD) = \frac{\xa(\GWa+4\pi\GWD)+\xb}{\xc(\GWa+4\pi\GWD)+\xd}\,.
$$
The solution is:
$$
\GWa = \frac{\xa-\xd}{2\xc}\,,\qquad
4\pi\GWD=\frac{\sqrt{(\xa+\xd)^2-4}}{2|\xc|}\,,
$$
and is real for a hyperbolic element of $\SL(2,\Z)$ (with $|\xa+\xd|>2$).
Note that it is important to have both $(a\pm 4\pi D)$ as fixed-points of the $\SL(2,\Z)$ transformation, so as not to reverse the orientation of the $\tau(x_3)$ curve, and not create a discontinuity in $\tau'(x_3)$.
So, given $\xa$, $\xb$, $\xc$, $\xd$, our configuration is constructed by first calculating $\GWa$ and $\GWD$, and then picking an arbitrary $\psi(2\pi)$ with a corresponding $\tau(2\pi)=\GWa+4\pi\GWD e^{2i\psi(2\pi)}$. Next, we calculate the $\SL(2,\Z)$ dual $\tau(0) = (\xa\tau(2\pi)+\xb)/(\xc\tau(2\pi)+\xd)$ and match it to a point on the semicircle according to $\tau(0)=\GWa+4\pi\GWD e^{2i\psi(0)}$. The function $\psi(x_3)$ can then be chosen arbitrarily, as long as it connects $\psi(0)$ to $\psi(2\pi)$. It can then be checked that $\GWr$ and $\GWtdr$ are continuous at $x_3=2\pi$.

At low-energy, the mass parameters $\GWr$ and $\GWtdr$ in $I'''$ make the scalar fields ($X^a$ and $Y^p$) massive.
Note that in principle, the parameters can be locally negative [although this can be averted by choosing $\psi(x_3)$ so that $\psi''=0$], but the effective 2+1D masses, [obtained by solving for the spectrum of the operators $-d^2/dx_3^2 + \GWr(x_3)$, and $-d^2/dx_3^2 + \GWtdr(x_3)$] have to be positive, since the configuration is supersymmetric and the BPS bound prevents us from having a profile of $X^a(x_3)$ or $Y^p(x_3)$ with negative energy. Similar statements hold for the fermionic masses in $I'$.

\subsection{The supersymmetry parameter}

As explained in \cite{Kapustin:2006pk}, the $\SL(2,\Z)$ duality transformation acts nontrivially on the SUSY generators.
Define the phase $\varphi$ by
$$
e^{i\varphi} = \frac{|\xc\tau+\xd|}{\xc\tau+\xd}\,.
$$
Then, the SUSY transformations act on the supersymmetry parameter as 
$$
\sParamEpsilon\rightarrow e^{\frac{1}{2}\varphi\Gamma_{0123}}\sParamEpsilon\,.
$$
(See equation (2.25) of \cite{Kapustin:2006pk}.)

We can now check that
\be\label{eqn:phasescancel}
\frac{|\xc\tau+\xd|}{\xc\tau+\xd} = e^{i(\tGWpsi-\GWpsi)}\,,
\ee
where $\tGWpsi$ is defined by
$$
\tilde{\tau}\equiv \frac{\xa\tau+\xb}{\xc\tau+\xd} \equiv\GWa + 4\pi\GWD e^{2 i\tGWpsi}\,.
$$
It follows from \eqref{eqn:phasescancel} that the Gaiotto-Witten phase that is picked up by the supersymmetry parameter as it traverses the Janus configuration from $\xI=0$ (corresponding to angular variable $\GWpsi$) to $\xI=2\pi$ (corresponding to $\tGWpsi$) is precisely canceled by the Kapustin-Witten phase of the $\SL(2,\Z)$-duality twist.
The entire ``Janus plus twist'' configuration is therefore supersymmetric.

\subsection{Extending to a type-IIA supersymmetric background}

In section \secref{sec:mapT} we assumed that there is a lift of the gauge theory construction to type-IIB string theory and, following a series of dualities, we obtained a type-IIA background with NSNS fields turned on.
Here we would like to outline how such a lift might be constructed.
We start with the well-known $AdS_3\times S^3\times T^4$ type-IIB background, and perform S-duality (if necessary) to get the $3$-form flux to be NSNS. Then, take $AdS_3$ to be of Euclidean signature and replace $T^4$ with $\R^4$, which we then Wick rotate to $\R^{1,3}$.
We take the $AdS_3$ metric in the form
$$
ds^2 = \frac{r^2}{r_1 r_5} (-dt^2 + dx_5^2)+\frac{r_1}{r_5}\sum_{i=6}^9 dx_i^2 + \frac{r_1 r_5}{r^2}dr^2
+ r_1 r_5 d\Omega_3^2
$$
$$
H^{(RR)} = \frac{2r_5^2}{g}(\epsilon_3 + {}_6^*\epsilon_3)\,,\qquad
e^\phi = \frac{g r_1}{r_5}
$$
where $\epsilon_3$ is the volume form on the unit sphere, and ${}_6^*$ is the Hodge dual in the six dimensions $x_0,\ldots,x_5$ (of $AdS_3\times S^3$), and where $r_1,r_5$ are constants.
(We follow the notation of
 \cite{Maldacena:1998bw}.)

We need to change variables $r\rightarrow x_3$, $t\rightarrow i x_1$ and $x_9\rightarrow i x_0$, and perform S-duality (where the RHS of arrows are the variables of \secref{sec:Duality}).
We then compactify directions $x_1$ and $x_2$ so that $0\le x_i<2\pi\xL_i$ ($i=1,2$).
As a function of $x_3$, we define the K\"ahler modulus of the $x_1-x_2$ torus to be
$$
\rho = i\frac{4\pi^2 r_1^2\xL_1\xL_2}{x_3^2} 
$$
Finally, we perform T-duality on direction $x_5$ to replace $\rho$ with the complex structure $\tau$ of the resulting $T^2$.
In an appropriate limit, this gives a solution where $\tau$ goes along a straight perpendicular line in the upper half plane.
We can convert it to a semi-circle with an $\SL(2,\R)$ transformation.

\end{appendix}


\bibliographystyle{my-h-elsevier}

\end{document}